\documentclass[conference]{IEEEtran}
\IEEEoverridecommandlockouts

\usepackage{mathrsfs}
\usepackage{amsfonts}
\usepackage{ifpdf}
\usepackage{cite}
\usepackage{array}
\usepackage{booktabs}
\usepackage{setspace}

\ifCLASSINFOpdf
\usepackage[pdftex]{graphicx}
\else \fi
\usepackage[cmex10]{amsmath}
\usepackage{array}
\usepackage{mdwmath}
\usepackage{mdwtab}
\usepackage{eqparbox}
\usepackage[tight,footnotesize]{subfigure}
\usepackage{algorithmic}
\usepackage{algorithm}
\usepackage{amsmath}
\usepackage{epsfig}
\usepackage{color}
\usepackage{ae}
\usepackage{amssymb}
\usepackage{times}
\usepackage{amsmath}   

\usepackage{xcolor}

\usepackage{amsthm} 

\usepackage{supertabular} 
\usepackage{stfloats}
\usepackage{multirow}

\newlength\savewidth

\begin{document}

\title{Channel Estimation for Underwater Acoustic OFDM Communications: An Image Super-Resolution Approach}
\author{
         Donghong Ouyang, Yuzhou Li, and Zhizhan Wang \\
         Huazhong University of Science and Technology, Wuhan, 430074, P. R. China\\
         State Key Laboratory of Integrated Services Networks (Xidian University), Xi'an, 710071, P. R. China\\
         \{dhouyang, yuzhouli, zhizhanwang\}@hust.edu.cn

\thanks{This work was supported in part by the National Science Foundation of China under Grant 61971462, Grant 61831013, and Grant 61631015 and the State Key Laboratory of Integrated Services Networks (Xidian University) under Grant ISN19-09.}
}
\maketitle
\IEEEpeerreviewmaketitle
\begin{abstract}
In this paper, by exploiting the powerful ability of deep learning, we devote to designing a well-performing and pilot-saving neural network for the channel estimation in underwater acoustic (UWA) orthogonal frequency division multiplexing (OFDM) communications. By considering the channel estimation problem as a matrix completion problem, we interestingly find it mathematically equivalent to the image super-resolution problem arising in the field of image processing. Hence, we attempt to make use of the very deep super-resolution neural network (VDSR), one of the most typical neural networks to solve the image super-resolution problem, to handle our problem. However, there still exist significant differences between these two problems, we thus elegantly modify the basic framework of the VDSR to design our channel estimation neural network, referred to as the channel super-resolution neural network (CSRNet). Moreover, instead of training an individual network for each considered signal-to-noise ratio (SNR), we obtain an unified network that works well for all SNRs with the help of transfer learning, thus substantially increasing the practicality of the CSRNet. Simulation results validate the superiority of the CSRNet against the existing least square (LS) and deep neural network (DNN) based algorithms in terms of the mean square error (MSE) and the bit error rate (BER). Specifically, compared with the LS algorithm, the CSRNet can reduce the BER by 44.74\% even using 50\% fewer pilots.
\end{abstract}

\section{Introduction}\label{Section:Introduction}
Orthogonal frequency division multiplexing (OFDM), a widely-used technology in terrestrial communication systems, has recently been employed in underwater acoustic (UWA) communications attributed to its efficient spectrum resource utilization and strong resistance to the multipath effect \cite{Acou_Summary_JOE2019}. However, unlike in terrestrial communication systems, it is not easy to exert the full potential of the OFDM in the extremely hostile UWA channel due to the violent path attenuation, strong doppler effect, and large ambient noise \cite{MarineMagazine_Mag2018,SurveyMI_Survey2019,ToRealyOrNot_TCOM2018}. Among all the techniques to overcome these challenges, accurate channel estimation plays a significant role in both the adaptive modulation at the transmitter and the signal detection at the receiver. Nevertheless, the pilot-based channel estimation methods, a kind of most widely-used approaches in terrestrial communication systems, are hard to be directly applied to UWA-OFDM systems, because the quite bandwidth-limited UWA channel makes it impossible to improve the estimation accuracy through increasing pilots as the terrestrial usually do \cite{Acou-bandwidth-limited_SIGMOBILE2006}. As a result, an important question that arises is how to achieve accurate UWA-OFDM channel estimation using as few pilots as possible.

There have been extensive works to investigate the UWA-OFDM channel estimation via different approaches, such as the least square (LS) \cite{LS&MMSE_VTC1995}, minimum mean-square error (MMSE) \cite{MMSE_Survey2007}, linear minimum mean-square error (LMMSE) \cite{LMMSE_TCOM1998}, and compressive sensing (CS) methods \cite{CS_MAP_GLOBCOM2018}. The LS, MMSE, and LMMSE \cite{LS&MMSE_VTC1995, MMSE_Survey2007, LMMSE_TCOM1998} are three kinds of the most
classic methods for terrestrial communication systems. By exploiting the sparsity of the UWA channel, the CS algorithm can be used to recover the original signal with a sampling rate lower than the Nyquist sampling rate. With this fact in mind, Mhd \textit{et al.} \cite{CS_MAP_GLOBCOM2018} combined the expectation maximization and the maximum posteriori probability methods to develop an iterative UWA channel estimation method under the assumption that UWA channels undergo Rayleigh fading.

Apart from the above approaches, deep-learning-based methods, a kind of methods that work well in the fields of computer vision, natural language processing, and so on, have recently also been employed for the channel estimation
problem. Hao \textit{et al.} \cite{ChannelEstimationImplictly_Jinshi2018} developed a deep neural network (DNN) based model to estimate the channel state information (CSI) implicitly without online training and experiment results verified its robustness when the number of training pilots is decreased. In \cite{UWA-OFDMChannelEstimation_Access2019}, two DNN-based models with different architectures were designed to solve the UWA-OFDM channel estimation problem and extensive experiments were performed to evaluate the performance of the proposed algorithm over the LS, MMSE, and back propagation algorithms.

To summarize, the LS, MMSE, and LMMSE algorithms proposed in \cite{LS&MMSE_VTC1995, MMSE_Survey2007, LMMSE_TCOM1998} would obtain unsatisfactory performance if being directly employed in UWA-OFDM systems as the characteristics of the UWA channels were not considered therein. Although \cite{CS_MAP_GLOBCOM2018} utilized the sparse feature of the UWA
channel to improve the estimation accuracy, the introduced iteration procedure would increase the computational complexity of the proposed algorithms. In spite of achieving remarkable estimation accuracy, \cite{UWA-OFDM Channel Estimation_Access2019} did not consider the impacts of the number of consumed pilots on the accuracy, which would unavoidably exacerbate the expenditure of spectrum resources originally for data transmission due to the extremely limited UWA bandwidth. Although \cite{ChannelEstimationImplictly_Jinshi2018} took these impacts into account, they trained a neural network for each considered signal-to-noise ratio (SNR), thus inevitably increasing application difficulties since the SNR cannot be known in advance.

In view of these, considering the outstanding performance of deep-learning-based methods, this paper devotes to designing a pilot-saving, high-performance, and unified neural network suitable for all the considered SNRs to accomplish UWA-OFDM channel estimation. To this end, we first analyze the essential attributes of the UWA channel and find this problem mathematically equivalent to the image super-resolution problem arising in the field of image processing. Thus, we make use of the very deep super-resolution convolutional network (VDSR) \cite{VDSR_CVPR2016}, one of the most typical neural networks to solve the image super-resolution problem, to design our channel estimation neural network. However, owing to the significant differences between these two problems, we modify the basic framework of the VDSR to match the characteristics of the UWA channel. Furthermore, instead of training an individual neural network for each SNR, we utilize transfer learning to obtain an unified network suitable for all the considered SNRs to increase the practicability of the proposed network. Simulation results show that the proposed neural network, referred to as the the channel super-resolution neural network (CSRNet), can achieve higher estimation accuracy even using much fewer pilots in the considered range of SNRs over some existing algorithms.

The remainder of this paper is organized as follows. In Section~\ref{Section:System Model}, we transform the channel estimation problem into the image super-resolution problem. Our proposed CSRNet is described in Section~\ref{Section:CsiNet}. Section~\ref{Section:Simultion Results} presents simulation results to evaluate the performance of the CSRNet. Finally, we conclude our paper in Section~\ref{Section:conclusion}.

\section{Problem Transformation}\label{Section:System Model}

\begin{figure}[t]
\centering \leavevmode \epsfxsize=3.5 in  \epsfbox{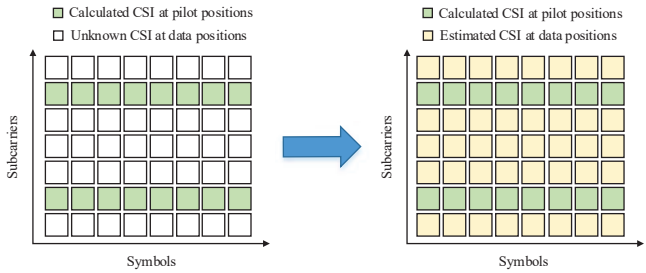}
\centering \caption{Illustration of the procedure of channel estimation. In this figure, the green cubes represent the CSI at pilot positions and the white and yellow ones represent the unknown and estimated CSI at data positions, respectively.} \label{Fig:channel estimation}
\end{figure}
It is known that the acquisition of the CSI plays a significant role in implementing the OFDM system as it is crucial for adaptive modulation and coding at the transmitter and accurate demodulation and recovery at the receiver. Among all the methods to obtain the CSI, the kind of pilot-based channel estimation methods is most commonly used in practical
communication systems. Specifically, as depicted in Fig.~\ref{Fig:channel estimation}, first calculate the CSI at pilot positions that are known at transmitters and receivers, and then estimate the CSI at data positions based on the calculated CSI at pilot positions.

From the above analysis and Fig.~\ref{Fig:channel estimation}, pilot-based channel estimation can be mathematically considered as a matrix completion problem, in which partially known elements in a CSI matrix are utilized to recover the remaining ones. Interestingly, we find that this problem is remarkably similar to the image super-resolution problem, a well-studied problem in the field of image processing. Specifically, a high-resolution image is obtained from its corresponding low-resolution image by recovering a complete image matrix from some partially known elements, and thus it also can be regarded as a matrix completion problem. Based on this fact, this paper attempts to make use of the solutions for image super-resolution problems to cope with our concerned channel estimation problem.

A vast number of image super-resolution techniques have been proposed in the literature, and among them, the interpolation-based method, the feature-space-construction-based method, and the deep-learning-based method are three typical ones. Furthermore, it has been verified that, compared with the first two methods, the deep-learning-based methods usually can achieve better performance in terms of both the recovery accuracy and the computational complexity in feature extraction \cite{SRCNN_PAMI2016}. Moreover, among all the state-of-the-art deep-learning-based methods of the image super-resolution, the very deep super-resolution convolutional network (VDSR) is one of the most representative ones. In particular, the VDSR outperforms the bicubic interpolation method by 3-4 dB and the anchored neighborhood regression by 1-2 dB in terms of the peak signal-to-noise ratio (PSNR)\cite{VDSR_CVPR2016}.

As described above, channel estimation and image super-resolution are mathematically equivalent and the VDSR works well for the image super-resolution problem. Based on these two facts, this paper adopts and modifies the basic framework of the VSDR to design our channel estimation neural network, which will be described in Section~\ref{Section:CsiNet}.

\section{Design of the Channel Estimation Neural Network for UWA-OFDM Communications} \label{Section:CsiNet}
Based on the above analysis, it can be found that the VDSR shows a great potential in accurately recovering the CSI matrix, nevertheless, the following two questions remain to be answered before deployment.
\begin{itemize}
\item Whether the VDSR could be directly applied to solve the UWA-OFDM channel estimation problem?
\item If not, how to modify the basic framework of the VDSR to match the features of the UWA-CSI matrix?
\end{itemize}

In this section, we first analyze the differences between the UWA-CSI and the image matrices, then elegantly modify the basic framework of the VDSR for the channel estimation problem, and finally develop the channel estimation neural network, referred to as the channel super-resolution neural network (CSRNet).

\begin{figure}[t]
\centering \leavevmode \epsfxsize=3.5 in  \epsfbox{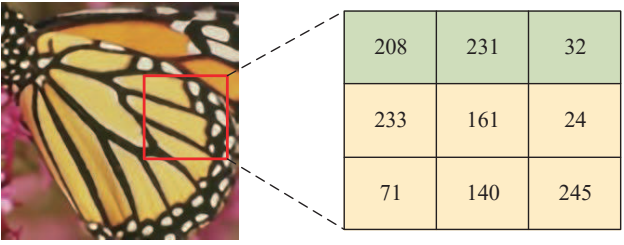}
\centering \caption{Example to show elements in the image matrix.} \label{Fig:elements in images}
\end{figure}

\begin{figure}[t]
\centering \leavevmode \epsfxsize=3.5 in  \epsfbox{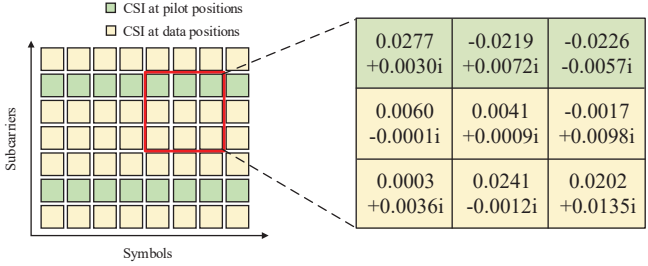}
\centering \caption{Example to show elements in the UWA-CSI matrix.} \label{Fig:elements in CSI}
\end{figure}

\subsection{Differences between Channel Estimation and Image Super-Resolution}\label{subsection:diffences}
Although channel estimation and image super-resolution can both be mathematically recast to the matrix completion problem, three significant differences, as depicted in  Figs.~\ref{Fig:elements in images} and ~\ref{Fig:elements in CSI}, exist between them, summarized as follows.

1) \emph{Complex vs real numbers:} In the image super-resolution problem, each image is usually represented by a three-channel RGB matrix and its elements are all real-valued. Differently, all elements in a UWA-CSI matrix are complex-valued, as depicted in Fig.~\ref{Fig:elements in CSI}.

2) \emph{Negative vs positive numbers:} Elements in an image matrix are all positive, as shown in Fig.~\ref{Fig:elements in images}, while elements in a UWA-CSI matrix contain both positive and negative numbers.

3) \emph{Quite small numbers vs integers:} Elements in an image matrix are all integers between $0$ and $256$. On the contrary, elements in a UWA-CSI matrix are quite small, usually at the order of $10^{-2}$ to $10^{-4}$, as shown in Fig~\ref{Fig:elements in CSI}.

From the above comparison, it can be obtained that the basic framework of the VDSR cannot be directly employed in the UWA-OFDM channel estimation, and thus necessary optimization and elegant modification are required.

\begin{figure}[t]
\centering \leavevmode \epsfxsize=3.5 in  \epsfbox{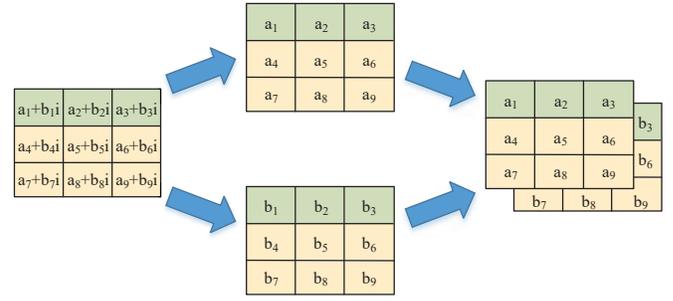}
\centering \caption{Procedure of separating the real and imaginary parts of the CSI matrix and reconstructing them as a new two-channel matrix.} \label{Fig:real_and_imag}
\end{figure}

\begin{figure*}[t]
\centering \leavevmode \epsfxsize=7.0 in  \epsfbox{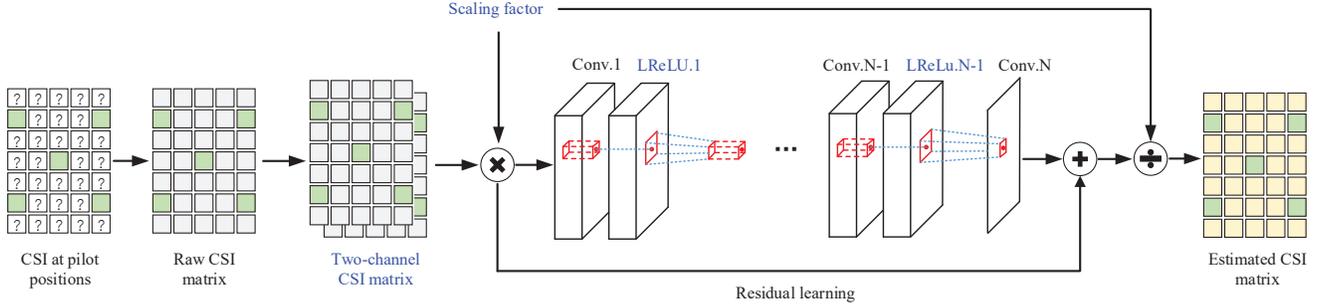}
\centering \caption{Architecture of our proposed CSRNet.} \label{Fig:CsiNet}
\end{figure*}

\subsection{Modification of the VDSR Framework for the CSRNet}\label{subsection:proposed improvement}
To maintain the advantages of the VDSR meanwhile overcome the above challenges imposed by the UWA-CSI matrix, this subsection modifies the basic framework of the VDSR for constructing the CSRNet from the following three aspects.

1) \emph{Separate real and imaginary parts into two channels:} To cope with the challenge that the VDSR cannot handle complex-valued numbers, we first separate each complex-valued UWA-CSI matrix into two matrices of real and imaginary parts as most neural networks do. By this, the problem of dealing with complex-valued numbers can be solved, but the underlying correlations between the real and imaginary parts are also discarded at the same time. In response to this problem, we then overlap and rebuild the two matrices as a new two-channel matrix, as shown in Fig.~\ref{Fig:real_and_imag}.

2) \emph{Select the LRelu as the activation function:} To handle the problem that negative numbers are not considered in the VDSR, we modify the rectified linear unit (ReLu) activation function used in the VDSR to guarantee the back propagation of negative numbers. To balance between an effective back propagation of negative numbers and a correct convergence for the developed network, we select the leaky rectified linear unit (LReLu) as our activation function, which is
\begin{equation}\label{Eq:LReLu}
\text{LReLu}(x)=
\begin{cases}
x, & x>0 \\
b_ix, & x\leq 0
\end{cases}
\end{equation}
where ${b_i}$ is a constant parameter between $0$ and $1$, typically set as $0.3$.

3) \emph{Magnify the CSI matrix with a scaling factor:} To deal with the vanishing gradient problem caused by quite small numbers, we introduce a scaling factor to preprocess the CSI matrix. After calculating the loss function (MSE) adopted in the VDSR, the values obtained from a CSI matrix will be very close to $0$, commonly at the order of $10^{-6}$, which is extremely likely to lead to the vanishing gradient. To avoid this problem, a feasible solution is to multiply the CSI matrix by a scaling factor before training and remove its impacts on the estimated CSI matrix by dividing the same factor after training.

\subsection{Proposed CSRNet Architecture}\label{subsection:CsiNet architecture}
Based on the above optimization and modification for the basic VDSR, we obtain the CSRNet architecture suitable for the UWA-OFDM channel estimation. Specifically, as shown in Fig.~\ref{Fig:CsiNet}, the CSRNet is composed of $20$ convolution layers, each of which is followed by the LReLu activation function. All the layers except the first and last ones are composed of $64$ filters with the same size of $3 \times 3 \times 64$. The first layer, i.e., the input layer, is comprised of $64$ filters with the size of $3 \times 3 \times m$, where $m$ represents the channel number of the input matrix. The last layer, i.e., the output layer, consists of $m$ filters with the size of $3 \times 3 \times 64$.

Besides, in order to avoid the vanishing gradient problem caused by very deep networks, we reserve the residual learning employed in the VDSR. Meanwhile, learning rate decay and early stopping strategies are adopted to accelerate the network convergence and avoid overfitting, respectively.

\subsection{Procedure of the Proposed Channel Estimation Method}\label{subsection:algorithm}
Based on the developed CSRNet, we now describe the whole channel estimation procedure, presented in Fig.~\ref{Fig:CsiNet}, from calculating the CSI at pilot positions to estimating all the elements in the CSI matrix.

\textit{Step 1}: Calculate the CSI at pilot positions. To reduce the computation complexity, we adopt the LS algorithm \cite{LS&MMSE_VTC1995}, a widely-used algorithm in practical communication systems, to calculate the CSI at pilot positions. However, the ambient noise has not been considered in the LS algorithm, thus its performance is usually unsatisfactory.

\textit{Step 2}: Obtain the raw CSI matrix. We estimate the remaining unknown elements at data positions from the surrounding calculated CSI at pilot positions through the spline interpolation method, a typical method in numerical analysis, to obtain the raw and inaccurate CSI matrix.

\textit{Step 3}: Transform the raw CSI matrix into the two-channel real-valued matrix. To match the complex-valued characteristics of the UWA-CSI matrix, we separate the real and imaginary parts of the each raw CSI matrix and then rebuild them as a two-channel real-valued CSI matrix.

\textit{Step 4}: Process the two-channel matrix with a scaling factor. To respond to the small-number property of the UWA-CSI matrix, a scaling factor is introduced to magnify the elements in the two-channel real-valued CSI matrix before training, whose impacts on the estimated CSI matrix will be removed by dividing the same factor after training.

\textit{Step 5}: Train the CSRNet. Taking the processed CSI matrix in Step 4 and the full-CSI matrix as the input and output of the CSRNet, respectively, we minimize the MSE as the loss function over the training set. Besides, to match the negative-number property of the UWA-CSI matrix, we choose the LRelu function as the activation function in the CSRNet.

\section{Simulation Results and Analysis}\label{Section:Simultion Results}
In this section, we first introduce the UWA parameter settings, and then present simulation results to evaluate the performance of the CSRNet in terms of the MSE and BER.

\subsection{Simulation Parameter Settings}\label{subection:parameters setting}
In our work, we adopt the widely-used UWA channel simulator developed in \cite{UWAchannelmodel_M2013}, which has been validated by realistic data obtained from four experiments, to generate the UWA channel, given by
\begin{equation}\label{Eq:H}
H(f,t)=\overline{H}_0\sum_p h_p\tilde{\gamma}_p(f,t)e^{-j2\pi f\tau_p}
\end{equation}
where ${\overline H _0}$, ${h_p}$, ${\tau_p}$, and $\tilde{\gamma}_p(f,t)$ are the nominal frequency response, large-scale path gains, delays and the small-scale fading of the $p$th propagation path. Specifically, the main simulation parameters for configuring the UWA physical environment and the UWA-OFDM system are summarized in Tables ~\ref{Table:physical environment} and ~\ref{Table:OFDM parameter settings}, respectively.

\begin{table} 
\centering
\normalsize
\caption{\label{Table:physical environment}Parameter settings of the considered UWA channel geometry and physical environment} 
\begin{tabular}{|p{3.5cm}<{\centering}|p{2.3cm}<{\centering}|} 
\hline
Parameters & Values \\
\hline
Water depth & 100 m \\
\hline
Transmitter depth  & 20 m \\
\hline
Receiver depth & 50 m \\
\hline
Transmission distance & 1 km \\
\hline
Spreading factor  & 1.7 \\
\hline
Sound speed in water & 1500 m/s \\
\hline
Sound speed in bottom & 1200 m/s \\
\hline
Number of intrapaths & 20 \\
\hline
Tx drifting speed & 0.1 m/s \\
\hline
Rx drifting speed & 0.02 m/s\\
\hline
Tx vehicular speed & $N(0,1)$ m/s\\
\hline
Rx vehicular speed & 0 \\
\hline
\end{tabular}
\end{table}

\begin{table} 
\centering
\normalsize
\caption{\label{Table:OFDM parameter settings}UWA-OFDM parameter settings} 
\begin{tabular}{|p{3.5cm}<{\centering}|p{1.3cm}<{\centering}|} 
\hline
Parameters & Values \\
\hline
Carrier frequency & 16 kHz \\
\hline
Channel bandwidth     & 4 kHz \\
\hline
No. of subcarriers & 512 \\
\hline
No. of symbols/frame & 16 \\
\hline
Subcarrier spacing & 7.81 Hz \\
\hline
Symbol duration & 128 ms \\
\hline
Modulation type & QPSK \\
\hline
\end{tabular}
\end{table}

Regarding the parameters of the CSRNet, we generate $10000$ CSI matrices, $80\%$ of which are randomly divided as the training set, $10\%$ of which are the validation set, and the last $10\%$ of which are the test set. Besides, we set the scaling factor to be $10$ and the initial learning rate to be $0.001$, which then attenuates by a factor of $0.1$ every $40$ epochs. The maximum epoch is set to be $100$ but training will early stop if the values of the loss function do not decline in $5$ continuous epochs.

As shown in Table ~\ref{Table:OFDM parameter settings}, the number of subcarriers is 512 and each frame is composed of 16 symbols, thus the size of a CSI matrix is 512 $\times$ 16. In our simulation, we adopt the LS algorithm \cite{LS&MMSE_VTC1995} and the DNN-based algorithm \cite{UWA-OFDMChannelEstimation_Access2019} to evaluate the performance of our proposed CSRNet, which are detailed as follows.

\begin{itemize}
\item \textit{LS-2 symbols:} In this configuration, 2 among 16 symbols, i.e., the 4th and 12th symbols, are first selected as pilots, and then the LS algorithm and the spline interpolation method are utilized to estimate the CSI at pilot and data positions, respectively.
\item \textit{LS-4 symbols:} In this configuration, it follows the same algorithm flow as that in the \textit{LS-2 symbols} but with 4 symbols, i.e., the 3th, 7th, 11th, and 15th symbols, being selected as pilots instead.
\item \textit{DNN-2 symbols:} In this configuration, as in the \textit{LS-2 symbols}, the 4th and 12th symbols are selected as pilots, and then the DNN-based algorithm is utilized to obtain the CSI matrix. Specifically, the DNN model in \cite{UWA-OFDMChannelEstimation_Access2019} is adopted and the neuron number in each layer is modified as 4, 64, 128, 64, and 32, respectively, to accommodate the above parameter settings.
\item \textit{DNN-4 symbols:} In this configuration, it follows the same algorithm flow as that in the \textit{DNN-2 symbols} but with 4 symbols, i.e., the 3th, 7th, 11th, and 15th symbols, being selected as pilots instead.
\item \textit{CSRNet-2 symbols:} In this configuration, as in the \textit{LS-2 symbols}, the 4th and 12th symbols are selected as pilots, and then the proposed CSRNet is utilized to obtain the CSI matrix.
\item \textit{CSRNet-4 symbols:} In this configuration, it follows the same algorithm flow as that in the \textit{CSRNet-2 symbols} but with 4 symbols, i.e., the 3th, 7th, 11th, and 15th symbols, being selected as pilots instead.
\item \textit{FullCsi:} In this configuration, the CSI generated by (\ref{Eq:H}) is utilized to recover the original data.
\end{itemize}

\subsection{Comparison between an Unified Network and Individual Networks for Different SNRs}\label{subsection:onenet}
\begin{figure}[t]
\centering \leavevmode \epsfxsize=3.5 in  \epsfbox{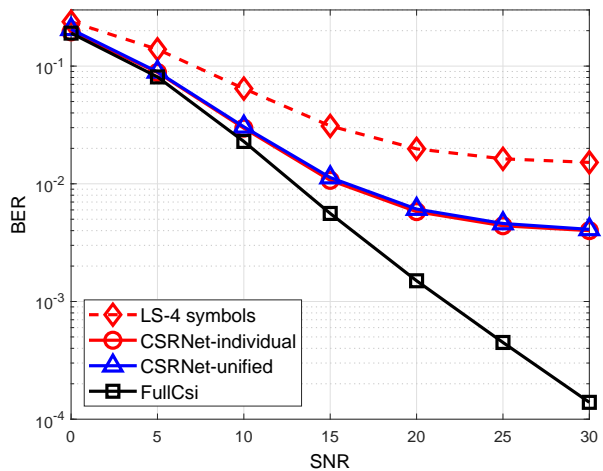}
\centering \caption{Comparison between the CSRNet-unified (a unified CSRNet for all considered SNRs) and the CSRNet-individual (each considered SNR associated with a CSRNet). In this figure, 4 symbols, i.e., the 3th, 7th, 11th, and 15th symbols, are selected as pilots for both the CSRNet-unified and CSRNet-individual.} \label{Fig:onenet}
\end{figure}

As a matter of fact, training an individual network for the each considered SNR \cite{ChannelEstimationImplictly_Jinshi2018} just satisfies the purpose of the validity but ignores the practicality of the proposed algorithm as the SNR usually cannot be known in advance. To obtain an unified network suitable for all the given SNRs, we utilize transfer learning, a commonly-used strategy in the field of machine learning, to accelerate the network convergence and improve performance. In general, transfer learning attempts to apply the knowledge obtained in one domain of interest to another similar domain, e.g., the features learned from classifying dogs may benefit th classification of cats. Based on this fact, we adopt the neural network obtained in case of SNR = $15$ dB as the pre-training network and freeze the first ten layers to train the network suitable for all the given SNRs. Fig.~\ref{Fig:onenet} shows that the performance of training an unified network is almost equal to that of training the individual network for each SNR, which significantly increases the practical value of the CSRNet.

\begin{figure}[t]
\centering \leavevmode \epsfxsize=3.5 in  \epsfbox{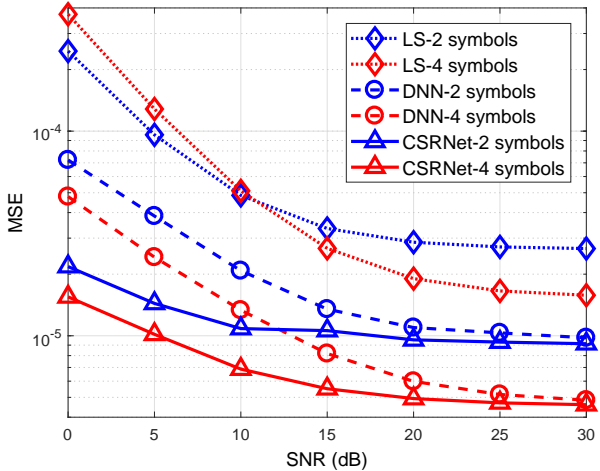}
\centering \caption{Comparison of the MSE versus SNR among the LS \cite{LS&MMSE_VTC1995}, the DNN-based \cite{UWA-OFDMChannelEstimation_Access2019}, and the CSRNet approaches under different pilot number settings.} \label{Fig:MSE}
\end{figure}

\subsection{Mean Square Error}\label{subsection:MSE}
In Fig.~\ref{Fig:MSE}, we evaluate the performance of the CSRNet against the LS and the DNN-based algorithms in terms of the MSE in case of SNR = $0-30$ dB, among which the DNN-based algorithm is considered to exhibit the comparable performance with the MMSE algorithm \cite{ChannelEstimationImplictly_Jinshi2018, UWA-OFDMChannelEstimation_Access2019}. In Fig.~\ref{Fig:MSE}, compared with the LS and the DNN-based algorithms, the MSEs of the CSRNet are dramatically reduced by $95.84\%$ and $67.64\%$ in case of SNR = $0$ dB, respectively.

Besides, we also show the impacts of the number of consumed pilots on the MSE in Fig.~\ref{Fig:MSE}. It can be found that the more pilots always bring the better performance except the LS algorithm in low SNRs. This result is ascribed to the fact that the elements in UWA-CSI matrices are averagely at the order of $10^{-4}$ after squared, while the MSEs in these situations are also at the same order. When the error is comparable with itself, it is not surprising that the MSE of anyone is better than the other because their performances are both disappointing.

\begin{figure}[t]
\centering \leavevmode \epsfxsize=3.5 in  \epsfbox{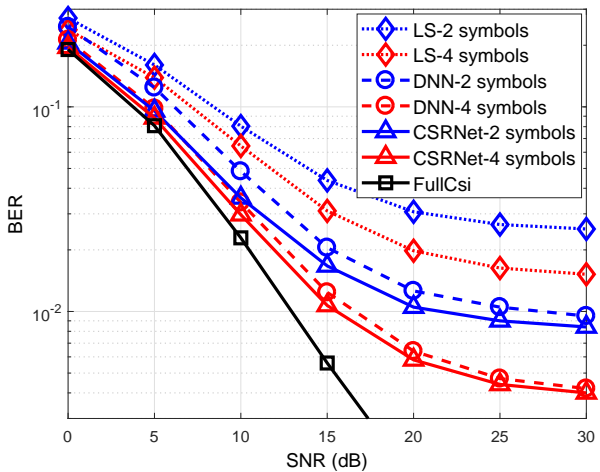}
\centering \caption{Comparison of the BER versus SNR among the LS \cite{LS&MMSE_VTC1995}, the DNN-based \cite{UWA-OFDMChannelEstimation_Access2019}, and the CSRNet approaches under different pilot number settings.} \label{Fig:BER}
\end{figure}

\subsection{Bit Error Rate}\label{subsection:BER}

Furthermore, Fig.~\ref{Fig:BER} compares the BERs to estimate the performance for data recovery of the CSRNet. It can be found that the BERs of the CSRNet are always lower than those of the LS and DNN-based algorithms for the given SNRs, for example, even up to $73.68\%$ when SNR = $30$ dB and $13.71\%$ when SNR = $20$ dB, respectively. More importantly,
the BERs of the CSRNet with 2-symbol pilots are lower than those of the LS algorithm with 4-symbol pilots, which implies that the CSRNet can maintain the more outstanding performance (BER reduction up to 44.74\%) even with much fewer pilots (equal to 50\%). This result is particularly meaningful for resource-scarce UWA-OFDM systems, as we can save more time-frequency resource for data transmission, thereby remarkably increasing the data rate.

\section{Conclusion} \label{Section:conclusion}
In this paper, we have devoted to designing a channel estimation algorithm for UWA-OFDM communication systems to achieve the high accuracy using as few pilots as possible with the help of deep learning. Specifically, we have interestingly found that the channel estimation problem and the image super-resolution problem can both mathematically
be regarded as the matrix completion problem. Then, owing to the significant differences between the CSI and the image
matrices, we have modified the basic framework of the VDSR, a typical neural network to solve the image super-resolution problem, to develop our proposed CSRNet. Most importantly, to increase the practicality of the CSRNet, we have utilized transfer learning to obtain an unified neural network suitable for all the considered SNRs rather than an individual network for each SNR as the SNR is hard to be known in advance. Extensive simulation results have verified that, compared with the LS algorithm, the CSRNet can reduce the BER by 44.74\% even using 50\% fewer pilots.
\bibliographystyle{IEEEtran}
\bibliography{IEEEabrv,CsiNet_Reference}

\end{document}